\documentclass[aps,showpacs,11pt,superscriptaddress,preprintnumbers,amsmath,amssymb,nofootinbib]{revtex4-1}
\pdfoutput=1
\usepackage{mathrsfs}
\usepackage{epsfig}
\usepackage{graphicx}
\usepackage{dcolumn}
\usepackage{bm}
\usepackage{amsmath}
\usepackage{slashed}       
\usepackage{amssymb}
\usepackage{amsfonts}

\let\jnfont=\rm
\def\NPB#1,{{\jnfont Nucl.\ Phys.\ B }{\bf #1},}
\def\PLB#1,{{\jnfont Phys.\ Lett.\ B }{\bf #1},}
\def\EPJC#1,{{\jnfont Eur.\ Phys.\ Jour.\ C }{\bf #1},}
\def\PRD#1,{{\jnfont Phys.\ Rev.\ D }{\bf #1},}
\def\PRL#1,{{\jnfont Phys.\ Rev.\ Lett.\ }{\bf #1},}
\def\MPLA#1,{{\jnfont Mod.\ Phys.\ Lett.\ A }{\bf #1},}
\def\JPG#1,{{\jnfont J.\ Phys.\ G}{\bf #1},}
\def\CTP#1,{{\jnfont Commun.\ Theor.\ Phys.\ }{\bf #1},}
\def\ZPC#1,{{\jnfont Z.\ Phys.\ C }{\bf #1},}
\def\JHEP#1,{{\jnfont JHEP \ }{\bf #1},}
\def\lsim{\raise0.3ex\hbox{$<$\kern-0.75em\raise-1.1ex\hbox{$\sim$}}}
\def\gsim{\raise0.3ex\hbox{$>$\kern-0.75em\raise-1.1ex\hbox{$\sim$}}}

\newcommand{\8}{\ \ \ \ \ \ \ \ }

\begin{document}

\preprint{IPMU15-0217}

\title{Interpreting the 750 GeV diphoton excess by the singlet extension \\ of the Manohar-Wise Model}
\author{Junjie Cao}
\email{junjiec@itp.ac.cn}
\affiliation{Department of Physics, Henan Normal University, Xinxiang 453007, China}
\author{Chengcheng Han}
\email{hancheng@itp.ac.cn}
\affiliation{Kavli IPMU (WPI), University of Tokyo, Kashiwa, Chiba 277-8583, Japan}
\author{Liangliang Shang}
\email{shlwell1988@gmail.com}
\affiliation{Department of Physics, Henan Normal University, Xinxiang 453007, China}
\author{Wei Su}
\email{weisv@itp.ac.cn}
\affiliation{State Key Laboratory of Theoretical Physics,
      Institute of Theoretical Physics, Academia Sinica, Beijing 100190, China}
\author{Jin Min Yang}
\email{jmyang@itp.ac.cn}
\affiliation{State Key Laboratory of Theoretical Physics,
      Institute of Theoretical Physics, Academia Sinica, Beijing 100190, China}
\author{Yang Zhang}
\email{zhangyang@itp.ac.cn}
\affiliation{State Key Laboratory of Theoretical Physics,
      Institute of Theoretical Physics, Academia Sinica, Beijing 100190, China}

\begin{abstract}
The evidence of a new scalar particle $X$ from the 750 GeV diphoton excess, and the absence of any other signal of new physics at the LHC so far
suggest the existence of new colored scalars,  which may be moderately light and thus can induce sizable $X g g$ and $X \gamma \gamma$ couplings without
resorting to very strong interactions. Motivated by this speculation, we extend the Manohar-Wise model by adding one gauge singlet scalar field.
The resulting theory then predicts one singlet dominated scalar $\phi$
as well as three kinds of color-octet scalars, which can mediate through loops the $\phi gg$ and $\phi \gamma \gamma$ interactions.
After fitting the model to the diphoton data at the LHC, we find that in reasonable parameter regions the excess can be explained at $1\sigma$ level
by the process $ g g \to \phi \to \gamma \gamma$, and the best points predict the central value of the excess rate with $\chi_{min}^2=2.32$, which corresponds to
a $p$-value of $0.68$. We also consider the constraints from
various LHC Run I signals, and we conclude that, although these constraints are powerful in excluding the parameter space of the model,
the best points are still experimentally allowed.

\end{abstract}

\maketitle

\section{Introduction}
Recently both ATLAS and CMS collaborations reported a resonance with its mass around 750 GeV in diphoton channel\cite{rrEX:ATLAS, rrEX:CMS},
and the local and global significances of this resonance are around $3.6\sigma$ and $2.3 \sigma$ respectively for the ATLAS analysis,
and $2.6 \sigma$ and $2 \sigma$ for the CMS analysis. Interestingly, as pointed out in \cite{Diphoton:UsedRate,new-analysis}
both the analyses favored the $750 {\rm GeV}$ diphoton production rate at about $4 {\rm fb}$ in the narrow-width approximation
\footnote{Currently with insufficient experimental data,
the ATLAS analysis slightly preferred a wide width of the resonance (about $45 {\rm GeV}$) to a narrow width \cite{rrEX:ATLAS}, and by contrast
the CMS analysis favored a narrow width \cite{rrEX:CMS}. Very recently, an analysis by combining both the ATLAS data and the CMS data
was carried out, and it indicated that the narrow width was preferred \cite{new-analysis}.}.
Obviously, if the resonance corresponds to a new particle beyond the Standard Model (SM), its spin should be either
$0$ or $2$. Since other searches for the $WW,ZZ,ll,jj$
signals at the LHC saw no excess at the resonance, the usual spin-2 KK-graviton, which has a universal coupling to SM particles \cite{Randall},
should be strongly limited by the resonant dilepton searches. So in the following we focus on the spin-0 case. Another aspect
we note is that if the resonance is initiated by $q\bar{q}$ annihilation, its production rate should be enhanced by a factor of $2.5$
from naive parton distribution analysis in moving from 8 TeV LHC to 13 TeV LHC \cite{HCS}. By contrast, if the resonance is initiated
by gluon fusion, the enhancement factor becomes 4.7. Given that the non-observation of the resonance at the 8 TeV
LHC has set an upper bound of about $2 {\rm fb}$ on the diphoton channel, it is better to consider gluon fusion channel
as the explanation of the excess.

As shown in previous studies \cite{Diphoton:UsedRate,DiphotonExcess}, if the new scalar $X$ only interacts with the SM
particles, it can not be fully responsible for the excess.  This is because both the $X g g$ and $X \gamma \gamma$ interactions
are induced by loop effects, and increasing the strength of the interactions will inevitably enhance the production rates
of the other SM particles, which has been tightly constrained by experimental data. So the diphoton excess implies the
existence of other new particles which contribute to the interactions of the $X$. So far additional fermions have been intensively
studied to enhance the diphoton production rate, but this kind of explanations usually need a rather strong Yukawa coupling and also a
somewhat low fermion mass scale, which suffers from rather tight theoretical and experimental constraints \cite{vacuum-stability}.
This motivates us to consider other types of new particles which act as the mediator to enhance the $X g g$ and
$X \gamma \gamma$ couplings. In this work, we take the singlet
extension of the Manohar-Wise model as an example to show that color-octet scalars are capable of doing such a work.
We note that although the color-octet scalars are well motivated in many basic theories, such as various SUSY constructions,
topcolor models and the models with extra dimensions \cite{color:model}, the attempt to interpret the excess
by such scalars is still absent.

This paper is organized as follows: in Sec.\ref{model}, we first introduce the singlet extension of the Manohar-Wise Model, then
in Sec.\ref{rule} we provide analytical formulae for the diphoton rate. The relevant constraints from the LHC Run I data
are described in Sec.\ref{const}, and our numerical results and discussions are presented in Sec.\ref{result}. For completeness,
we also discuss some theoretical issues of our explanation in subsequent section. Finally, we draw our conclusions in Sec.\ref{sum}.

\section{The Singlet extension of the Manohar-Wise Model}\label{model}
Motivated by the principle of minimal flavor violation,
the Manohar-Wise model extends the SM by one family scalars in the $(8,2)_{1/2}$ representation of the SM gauge group
$SU(3) \bigotimes SU(2) \bigotimes U(1)$ \cite{cos-Model}.
These scalars can be written as
\begin{eqnarray}
S^{A}=\left( \begin{array}{c}
        S^A_+ \\
        \frac{1}{\sqrt{2}} (S^A_R+ i S^A_I)
      \end{array} \right),
\end{eqnarray}
where $A=1,...,8$ is color index, $S^A_+$ denotes an electric charged color-octet scalar field,
and $S^A_{R}$ and $S^A_{I}$ are neutral CP-even and CP-odd ones respectively. In order to
explain the diphoton excess, we further incorporate one real gauge singlet scalar field $\Phi$ into the theory.
Then the general renormalizable scalar potential is given by
\begin{eqnarray}
V&=& m_\Phi^2 \Phi^2 + \lambda_\Phi \Phi^4 + m_H^2 H^\dagger\cdot H + \lambda_H (H^\dagger\cdot H)^2 +
\lambda_{H \Phi} \Phi^2 H^\dagger\cdot H \nonumber \\
&& + \ 2 m_{8}^2 \ \text{Tr} (S^{\dagger i}S_i) + \kappa \Phi^2 \text{Tr} (S^{\dagger j}S_j)  +
\lambda_1 H^{\dagger
i}H_i \text{Tr} (S^{\dagger j}S_j) + \lambda_2 H^{\dagger i}H_j
\text{Tr} (S^{\dagger j}S_i)
\nonumber\\
&& + \big [ \lambda_3 H^{\dagger i}
H^{\dagger j} \text{Tr}(S_iS_j) + \lambda_4 H^{\dagger i}
\text{Tr}(S^{\dagger j}S_j S_i) +  \lambda_5 H^{\dagger i}
\text{Tr}(S^{\dagger j}S_i S_j) + h.c. \big ]
\nonumber\\
&& + \lambda_6 \text{Tr}
(S^{\dagger i}S_i S^{\dagger j} S_j) + \lambda_7
\text{Tr}(S^{\dagger i}S_j S^{\dagger j} S_i) + \lambda_8 \text{Tr}
(S^{\dagger i}S_i)\text{Tr}(S^{\dagger j}S_j)\8 \8 \nonumber\\
&& + \lambda_9 \text{Tr}
(S^{\dagger i} S_j) \text{Tr}(S^{\dagger j}S_i) + \lambda_{10}
\text{Tr} (S_i S_j) \text{Tr}(S^{\dagger i}S^{\dagger j})
+\lambda_{11} \text{Tr}(S_iS_j S^{\dagger j} S^{\dagger i}),  \label{Lag}
\end{eqnarray}
where $S$ is the sum of the product $S^A T^A$ over the color index $A$,  $H=\frac{1}{\sqrt{2}} U (0, H^0)^T$ stands for the SM Higgs field
in unitary gauge, $i,j$ are isospin indices, and the dimensionless coefficients $\lambda_\Phi$, $\lambda_H$, $\lambda_{H \Phi}$, $\kappa$ and
$\lambda_\alpha$ ($\alpha=1,..., 11$) parameterize the interactions among the scalar fields. Note that all these coefficients
except $\lambda_4$ and $\lambda_5$ are real parameters \cite{cos-Model}, and their magnitudes may reach $20$ without conflicting with the
unitarity constraint \cite{Cao-Octet}.

Within this theoretical framework, two CP-even color-singlet scalar particles $\phi$ and $h$,
and three kinds of color-octet scalar particles $S_+^A$, $S_R^A$ and $S_I^A$ are predicted. In the basis $(H^0, \Phi)$, the squared mass matrix
of the color-singlet fields is given by
\begin{eqnarray}
\left( \begin{array}{cc}
  m_H^2 + 3 \lambda_H v^2 +  \lambda_{H\Phi} f^2  &  \lambda_{H\Phi} v f  \\
   \lambda_{H\Phi} v f  & 2 m_\Phi^2 + 12 \lambda_\Phi f^2 +  \lambda_{H\Phi} v^2 			 \\
\end{array} \right),
\end{eqnarray}
where $v$ and $f$ are the vacuum expectation values (vev) of the fields $H^0$ and $\Phi$ respectively. After diagonalizing this matrix,
we have
\begin{eqnarray}
h &=& H^0 \cos \theta  +  \Phi \sin \theta,  \\
\phi & = & - H^0 \sin \theta  + \Phi \cos \theta,
\end{eqnarray}
with $\theta$ parameterizing the mixing of the fields. In our scheme for the excess, $h$ corresponds to the SM-like Higgs boson
discovered at the LHC, and $\phi$ is responsible for the diphoton signal by the process $g g \to \phi \to \gamma \gamma$.
So in the following, we set $m_h = 125 {\rm GeV}$, $m_\phi = 750 {\rm GeV}$ and $v = 246 {\rm GeV}$, and for the
convenience of discussion, we choose $f$ and $\sin \theta$ as the input parameters of the model. In this way,  we have following relationships
\begin{eqnarray}
\lambda_{H\Phi} &=& \frac{(m_h^2 - m_\phi^2 ) \sin \theta \cos \theta}{v f},  \\
\lambda_H &=& \frac{m_h^2 \cos^2 \theta + m_\phi^2 \sin^2 \theta }{2 v^2},  \\
\lambda_\Phi &=& \frac{m_h^2 \sin^2 \theta + m_\phi^2 \cos^2 \theta }{8 f^2}. \label{parameters}
\end{eqnarray}

As for the color-octet particles, their masses are given by
\begin{eqnarray}
m_{S_+}^2 &=& m_8^2 + \kappa f^2 + \lambda_1 \frac{v^2}{4},   \\
m_{S_R}^2 &=& m_8^2 + \kappa f^2 + (\lambda_1 + \lambda_2 + 2\lambda_3) \frac{v^2}{4}, \\
m_{S_I}^2 &=& m_8^2 + \kappa f^2 + (\lambda_1 + \lambda_2 - 2\lambda_3) \frac{v^2}{4},
\end{eqnarray}
and the coefficients of their interactions with the color-singlet particles are given by
\begin{eqnarray}
g_{hS^{A\ast}_i S^B_i} &\equiv& \tilde{Y}_i \delta^{AB} = (\frac{v}{2} \lambda_i \cos \theta + \kappa_i f \sin \theta ) \delta^{AB}, \nonumber \\
g_{\phi S^{A\ast}_i S^B_i} &\equiv& Y_i \delta^{AB} = (- \frac{v}{2} \lambda_i \sin \theta + \kappa_i f \cos \theta )  \delta^{AB}, \label{phi-coupling}
\end{eqnarray}
where $i=+, R, I$, and we define $\lambda_+ = \lambda_1$,
$\lambda_{R,I}=\frac{1}{2}(\lambda_1+\lambda_2 \pm 2 \lambda_3)$, $\kappa_+ = \kappa$, and
$\kappa_R = \kappa_I = \frac{\kappa}{2}$. Throughout this work, just for simplicity we set $\lambda_{1-11}$ equal to $0.1$
so that  all the colored scalars are nearly degenerate (we label their common mass by $m_S$ hereafter).
This assumption together with the requirements $|\sin \theta| \lesssim 0.01$ and $m_S > 500 {\rm GeV}$
(see discussion below) imply that the $h\gamma \gamma$ and $h g g$ couplings are only slightly changed
by the $S_i$-mediated loops, which is actually favored by the $125 {\rm GeV}$ Higgs data \cite{Cao-Octet}.
Note that in such a case the $S$, $T$ and $U$ variables are scarcely changed \cite{EWPD}.  Also note that the unbroken of the electric charge and
color symmetries requires the vevs of the colored scalars to vanish, and the vacuum stability at the scale $m_\phi = 750 {\rm GeV}$ requires \footnote{ The vacuum
stability at the electroweak scale was widely discussed in the extension of the SM by scalar fields, see for example the appendix A in \cite{octet-vacuum}.}
\begin{eqnarray}
m_{S}^2 > 0, \quad \quad 4 \lambda_H \lambda_\Phi - \lambda_{H \Phi}^2 > 0.
\end{eqnarray}
Obviously, these requirements are satisfied in our scheme.

In this work, we also assume the Yukawa couplings of the $S_i$ with quarks are negligibly small so that the $S_i$ will decay mainly through
the loops mediated by the colored scalars \cite{S-decay1}. In this case, $S_{R, I}$ may decay into $g g$, $g Z$ and $g \gamma$ with the $g g$ mode being the
dominant one \cite{S-decay2}. We checked that for $|\lambda_4|, |\lambda_5| \sim {\cal{O}}(0.1)$ and $m_{S} \simeq 600 {\rm GeV}$, the widths of $S_{R,I}$
are at the order of $10^{-3} {\rm MeV}$.

\section{Theoretical prediction of the diphoton rate } \label{rule}

\begin{figure}[t]
\centering
\includegraphics[width=12cm]{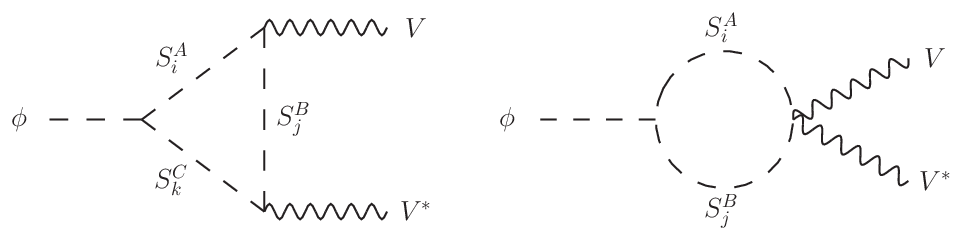}
\vspace*{-0.5cm}
\caption{Feynman diagrams contributing to the $\phi V V^\ast$ interactions with $VV$ denoting any of $\gamma \gamma$, $g g$, $Z \gamma$, $ZZ$ and $WW^\ast$.}
\label{fig1}
\end{figure}

In the extension of the Manohar-Wise model, the singlet dominated scalar $\phi$ can couple to vector boson pairs through its
mixing with the SM Higgs field $H^0$ or through the loop diagrams shown in Fig.\ref{fig1}.  As a result, $\phi$ may decay into
$\gamma \gamma$, $g g$, $ZZ$, $WW^\ast$ and $Z \gamma$. In the following we list the partial widths of $\phi$, which are needed to get the diphoton production rate.
\begin{itemize}
\item The widths of $\phi \to \gamma \gamma, g g, Z \gamma$ are given by
\begin{eqnarray}
  \Gamma_{\phi\to \gamma\gamma} & = & \frac{G_\mu \alpha^2 m_\phi^3}{128\sqrt{2}\pi^3}
  \left|\frac{4 Y_+ v}{m_{S_{+}}^2}A_0(\tau_{S_+}) - \sin \theta \times (A_1(\tau_W)
  +\frac{4}{3}A_{\frac{1}{2}}(\tau_t) ) \right|^2,  \\
  \Gamma_{\phi\to gg}&=& \frac{G_\mu \alpha_s^2 m_\phi^3}{16\sqrt{2}\pi^3}
  \left|\sum\limits_{i= +,R,I}\frac{3 Y_i v}{2 m_{S_i}^2} A_0(\tau_{S_i})
  - \frac{\sin \theta}{2}A_{\frac{1}{2}}(\tau_t)\right|^2,  \\
  \Gamma_{\phi\to Z\gamma} & = & \frac{G_\mu^2 m_W^2 \alpha m_\phi^3}{64\pi^4}
  \left( 1-\frac{m_Z^2}{m_\phi^2} \right)^3
  \left| \frac{4 Y_+ v}{m_{S_{+}}^2} \frac{1-2\sin^2\theta_W}{\cos\theta_W}C_0(\tau_{S_+}^{-1}, \eta_{S_+}^{-1}) \right. \nonumber \\
  & & \left.  - \sin \theta \times ( \cos\theta_W C_1(\tau_W^{-1},\eta_W^{-1})
  +\frac{2(1-\frac{8}{3}\sin^2\theta_W)}{\cos \theta_W} C_{\frac{1}{2}}(\tau_t^{-1},\eta_t^{-1}) )\right|^2,  \label{expression-gg}
\end{eqnarray}
where $A_0$, $A_{\frac{1}{2}}$, $A_1$, $C_0$, $C_{\frac{1}{2}}$ and $C_1$ are loop functions defined in \cite{h-rev}
with $\tau_\xi = m_\phi^2/(4 m_{\xi}^{2})$ and $\eta_\xi = m_Z^2/(4 m_{\xi}^{2})$ ($\xi = W$, $t$, $S_i$).
Note that in these expressions, terms proportional to $\sin \theta$ come from the
$H^0$-component of $\phi$, which can couple directly to top quark and W boson.
Also note that in the case of a small $\sin \theta $, we have
$\Gamma (\phi \to \gamma \gamma):\Gamma (\phi \to Z \gamma):\Gamma (\phi \to g g) \simeq 1: \frac{\sin^2 2 \theta_W}{2 \cos^2 2 \theta_W}: \frac{9}{2} \frac{\alpha_s^2}{\alpha^2} \simeq 1: 0.85: 715$ if the colored scalars are degenerated in mass.

\item The calculation of $\Gamma (\phi \to V V^\ast )$ with $V = W, Z$ is slightly complicated.
On one side, the $\phi W W^\ast$ and $\phi Z Z$ couplings have tree level contributions from the $H^0$-component of $\phi$,
which are proportional to $\sin\theta$, and thus suppressed if $\sin \theta \sim 0$. On the other side, the
couplings may get radiative corrections from the $S_i$-mediated loops, which might be sizable for a large $\kappa f$
and moderately light colored scalars. So for the sake of completeness, we calculate the both contributions.
In more detail, we first parameterize the effective $\phi V V^\ast$ interaction as
\begin{eqnarray}
{\cal{A}}_{\phi V V^\ast} = g_V m_V ( A_V g^{\mu \nu} + B_V p_2^{\mu} p_1^{\nu} ) \epsilon_\mu (p_1) \epsilon_{\nu}^\ast (p_2),
\end{eqnarray}
and express the decay width of $\phi \to V V^\ast $ by\cite{VV-amplitude}
\begin{eqnarray}
\Gamma_{\phi \to V V^\ast} &= & \delta_V
\frac{G_F M_\phi^3}{16 \pi \sqrt{2}} \frac{4 m_V^4}{m_\phi^4} \sqrt{\lambda(m_V^2,m_V^2;m_\phi^2)} \times \nonumber \\
&& \left[ A_V A_V^\ast \times \left ( 2 + \frac{(p_1 \cdot p_2)^2}{m_V^4} \right ) + ( A_V B_V^\ast + A_V^\ast B_V) \times
\left ( \frac{(p_1 \cdot p_2)^3}{m_V^4} - p_1 \cdot p_2 \right )  \right . \nonumber \\
&& \left. \ \ +\  B_V B_V^\ast \times \left ( m_V^4 + \frac{(p_1 \cdot p_2)^4}{m_V^4} - 2 (p_1 \cdot p_2)^2 \right ) \right ],
\end{eqnarray}
where $\delta_V=2(1)$ for $V=W(Z)$ respectively and $\lambda(x,y,z)= ((z-x-y)^2 - 4 xy)/z^2$. Then we compute the coefficients
$A_V$ and $B_V$ up to one loop level, which are given by
\begin{eqnarray}
  \nonumber
  A_V &=& -\sin\theta+ \frac{1}{2 \pi^2 v}\left(C_{1V} Y_+ + Y_R + Y_I \right)
  \left(B_0- 4C_{24}\right), \\
  \nonumber
  B_V &=& -\frac{1}{2 \pi^2 v}\left(C_{1V} Y_+ + Y_R + Y_I \right)
  \left(4C_{12}+4C_{23}\right).
\end{eqnarray}
In above expressions, $C_{1V} = 1, \cos^2 2 \theta_W$ for $V=W, Z$ respectively, the couplings $Y_i$
are defined in Eq.(\ref{phi-coupling}), and $B_0$, $C_{24}$, $C_{12}$, and $C_{23}$ are all loop functions defined in \cite{loop-definition}
with their dependence on external vector boson momenta and internal particle masses given by $B(-p_1-p_2,m_{S},m_{S})$ and $C(-p_1,-p_2,m_{S},m_{S},m_{S})$.

\item The width of $\phi \to t \bar{t}$ is given by
\begin{eqnarray}
\Gamma_{\phi \to t\bar{t}} &=& \sin^2 \theta \frac{3G_\mu}{4\sqrt{2}\pi} m_{\phi} m_t^2 \left(1-\frac{4m_t^2}{m_\phi^2}\right)^\frac{3}{2}.
\end{eqnarray}
Note that since we have neglected the Yukawa couplings of the $S_i$ with quarks, the $\phi \bar{t} t$ interaction is solely
determined by the $H^0$-component of $\phi$.
As a result, $\Gamma_{\phi \to t \bar{t}}$ as the largest one among all $\phi \to f \bar{f}$ decays is proportional to $\sin^2 \theta$.

\item The width of $\phi \to h h$ is given by
\begin{eqnarray}
\Gamma_{\phi \to h h} &=& \frac{\left|C_{\phi hh}\right|^2}{4\pi m_{\phi}^2}\left(\frac{m_\phi^2}{4}-m_h^2\right)^\frac{1}{2},
\end{eqnarray}
where
\begin{eqnarray}
C_{\phi h h} &=& - 3 \lambda_H v \sin\theta\cos^2\theta + 12 \lambda_\Phi f \sin^2\theta\cos\theta \nonumber \\
          && + \lambda_{H \Phi} (-v\sin^3\theta+f\cos^3\theta-2f\sin^2\theta \cos\theta+2v \sin\theta \cos^2\theta) \nonumber \\
          &\simeq & -\frac{m_\phi^2}{v} \sin \theta.
\end{eqnarray}
Note that in getting the final expression of $C_{\phi hh}$, we only keep the terms proportional to $\sin \theta$, and drop
higher order contributions in the expansion of $\sin \theta$.
\end{itemize}

The total width of $\phi$ is then given by
\begin{eqnarray}
\Gamma_{tot} = \Gamma_{\phi \to g g} + \Gamma_{\phi \to \gamma \gamma} + \Gamma_{\phi \to Z \gamma} + \Gamma_{\phi \to Z Z} + \Gamma_{\phi \to W W^\ast} + \Gamma_{\phi \to f \bar{f}} + \Gamma_{\phi \to h h} + \Gamma_{new},
\end{eqnarray}
where $\Gamma_{new}$ represents the width for other decay modes of $\phi$. These modes may arise if the theory is embedded in a more complex theoretical framework.

With these formula, the $\phi$-induced diphoton rate can be written as
\begin{eqnarray}
\sigma_{\gamma \gamma}^{13 TeV} = \frac{\Gamma_{\phi \to gg}}{\Gamma^{SM}_{H \to g g}} |_{m_H \simeq 750 {\rm GeV}} \times \sigma^{SM}_{\sqrt{s}=13 {\rm TeV}} (H) \times
\frac{\Gamma_{\phi \to \gamma \gamma}}{\Gamma_{tot}},
\end{eqnarray}
where $\Gamma^{SM}_{H \to g g}$ denotes the decay width of the SM Higgs $H$ into $g g$ with $m_{H} = 750 {\rm GeV}$, and $\sigma^{SM}_{\sqrt{s}=13 {\rm TeV}} (H)=735 {\rm fb}$  is the NNLO production rate of $H$ at the 13 TeV LHC \cite{HCS}. This expression indicates that if $\Gamma_{tot} \simeq \Gamma_{\phi \to gg}$, we have $\sigma_{\gamma \gamma}^{13 TeV} \varpropto (\kappa f)^2$;
and in comparison if  $\Gamma_{tot}$ is fixed at a certain value, $\sigma_{\gamma \gamma}^{13 TeV} \varpropto (\kappa f)^4$.

\section{Experimental constraints} \label{const}

Beside the diphoton signal, our model also predicts some other signals like $jj$, $VV^\ast$, $hh$ and $t\bar{t}$. Given the corresponding searches
done by the ATLAS and CMS collaborations with about $20 fb^{-1}$ data at the LHC Run I, one can naturally ask if these signals can be used to limit
our explanation. In the following, we briefly describe these searches.
\begin{itemize}
\item Resonant dijet signal.

 The dijet events at the 8 TeV LHC were studied in \cite{jj:ATLAS} and \cite{jj:CMS} by ATLAS and CMS respectively.
 ATLAS  provided an observed 95\% C.L. upper limit of  $11 \  {\rm pb}$ and $15 \  {\rm pb}$ on the rate for a $m = 750\ {\rm GeV}$ Breit-Wigner
 resonance with $\Gamma / m = 0.5\%$ and $ 5\%$ respectively (see Fig.9 in  \cite{jj:ATLAS} ), given that  the resonance is initiated by gluon fusion. By contrast,
 CMS imposed a much lower upper limit, which is about $1.8 \  {\rm pb}$  at 95\% C.L. for a $750 {\rm GeV}$ narrow
 resonance decaying into the $gg$ final state (see Fig.3 in  \cite{jj:CMS} ).

\item Resonant $hh$ signal.

CMS searched for the resonant SM-like Higgs pair production by the four bottom quark signal, and it found that $\sigma(pp\to X \to h h \to b
\bar{b} b \bar{b}) \leq 17$ fb for a spin-0 resonance with $m_{X}=750$ GeV (see Fig.5 in \cite{hh:CMS}). More comprehensive analyses for the resonant production
were done by ATLAS, focusing on the $b b \tau\tau$ and $\gamma\gamma W W^*$ final states \cite{hh:ATLAS:1}, the $\gamma\gamma b \bar{b}$
final state \cite{hh:ATLAS:2} and also the $b \bar{b}b \bar{b}$ final state \cite{hh:ATLAS:3}. Especially, the results of
the $b \bar{b}b \bar{b}$ and $ b b \tau\tau$ analyses were combined for $m_X > 500$ GeV, and
a 95\% upper limit of 35 fb on $\sigma(gg\to X \to hh)$ was obtained for $m_X = 750 {\rm GeV}$ (see Fig.6 in \cite{hh:ATLAS:1}).

\item Resonant $VV^\ast$ signal.

Based on up to 5.1 fb$^{-1}$ data at the 7 ${\rm TeV}$ LHC and up to 19.7 fb$^{-1}$ data at the 8 ${\rm TeV}$ LHC, CMS combined results from the search for a heavy Higgs boson $H$ by $l\nu l\nu$ and $l\nu qq$ final states from $H \to WW^\ast$ decay channel, and also from the search by $ll\nu \nu$, $llll$, $ll\tau\tau$ and $llqq$ final states from $H \to ZZ$ decay \cite{VV:CMS}. In either case, it obtained a 95\% C.L. upper limit of about $83 \  {\rm fb}$ for the cross section $\sigma(pp\to H \to V V^\ast)$ at the
8 TeV LHC with $m_H = 750 {\rm GeV}$ (see Fig.7 in \cite{VV:CMS}). In parallel, ATLAS performed similar studies, and it obtained
following $95\%$ C.L. upper limits for $m_H = 750 {\rm GeV}$: $\sigma(gg\to H \to WW^\ast) \leq 54\  {\rm fb}$, $37\ {\rm fb}$ in the
complex-pole scheme (where the width increases with $m_H$) and  the narrow-width
approximation respectively (see Fig.12 and  Fig.13 in \cite{WW:ATLAS}), and $\sigma(gg\to H \to ZZ) \leq 12\  {\rm fb}$ (see Fig.12 in \cite{ZZ:ATLAS}).

\item Resonant $Z \gamma$ signal.

ATLAS searched for the process $p p \to \phi \to Z\gamma \to \bar{l} l \gamma$\ \cite{Zr:ATLAS} with $l$ denoting either $e$ or $\mu$, and at the resonance $m_\phi = 750 {\rm GeV}$, it obtained an upper bound on the production rate at $0.24 \ {\rm fb}$ when $\phi$ is a pseudo-Goldstone boson and at $0.31$ fb when $\phi$ is a technimesons (see Fig.3 in \cite{Zr:ATLAS}).

\item Resonant $t\bar{t}$ signal.

CMS performed a search for the production of a heavy resonance decaying into $t\bar{t}$ \cite{tt:CMS}, and it set $95\%$ C.L. upper limits
on $\sigma(pp\to X \to t\bar{t})$ for the resonance at $750 {\rm GeV}$, which were $450 \ {\rm fb}$ and $550\ {\rm fb}$ for $X$ corresponding to a
vector boson $Z'$  in the narrow width case and the wide width case ($\Gamma/m=10\%$) respectively, and $700\  {\rm fb}$ for $X$ corresponding to a KK gluon in the Randall-Sundrum model
(see Fig.14 in \cite{tt:CMS}).  ATLAS used lepton-plus-jets events to search for the $t\bar{t}$ resonances, and it excluded
the narrow spin-0 scalar resonance at $750 {\rm GeV}$ with the production cross section greater than $700 \ {\rm fb}$ (see Fig.11 in \cite{tt:ATLAS}).

\end{itemize}

\begin{table}[t]
\centering
\caption{The tightest limits on various $750 {\rm GeV}$ resonant signals at the 8 TeV LHC set by either ATLAS or CMS collaboration. \label{tab1}}
\begin{tabular}{lcccccccccc}
\hline
\hline
~~Decay mode:            & $jj$          & $hh$    & $WW^\ast$       & $ZZ$      & Z$\gamma$ & $t\bar{t}$  \\
\hline
~~95\% C.L. limit:   ~~~~~& ~~~1800 fb~~~   & ~~~35 fb~~~    & ~~~37 fb~~~~     & ~~~12 fb~~~   & ~~~3.6 fb~~~            &  ~~~450 fb~~~    \\
\hline
\end{tabular}
\end{table}

From above discussion, one can get the tightest limits on various channels at the $750 {\rm GeV}$ resonance, which are summarized in Table \ref{tab1}.
In the following, we will use these limits to select the parameter space of the model.  As for the diphoton signal at the 8 TeV LHC, we note that only
mild upward fluctuations at the mass window around 750 GeV were seen by  ATLAS \cite{rr:ATLAS} and CMS \cite{rr:CMS}, and they are actually consistent with
the observed resonance at the 13 TeV LHC after considering the large statistical fluctuation. So to treat the diphoton signal in an unprejudiced way, we combine
the diphoton data at both the 8 TeV and the 13 TeV LHC together to fit our model. In practice, we adopt the method in \cite{Diphoton:UsedRate} to do such a work.
The data we take are \footnote{Note that in \cite{Diphoton:UsedRate}, the experimental data were extracted
from the 95\% C.L. expected and observed exclusion limits of the diphoton rate published by ATLAS and CMS. For the
$13\  {\rm TeV}$ ATLAS data, the authors assumed that they obey Poissonian distribution
to account for the large difference between the observed limit and the expected one, while for the rest data,
the authors assumed that they are Gaussian distributed. When we reproduced the Fig.1 of \cite{Diphoton:UsedRate}, we noted that the authors might
have used $0.63\pm 0.35 ~{\rm fb}$, instead of $0.63\pm 0.25 ~{\rm fb}$ presented in the text of \cite{Diphoton:UsedRate},
as the CMS $8\ {\rm TeV}$ data in performing the fit.
Anyhow, we checked that the two choices of the CMS data do not result in significant difference about the main conclusions of this work. }
\begin{eqnarray}\label{chisq}
\mu^{exp}_i=\sigma(pp\to \gamma\gamma)=\left\{\begin{array}{lllll}
0.63\pm 0.25 ~{\rm fb} 	& ~~~{\rm CMS}    &{\rm at} ~\sqrt{s} = 8 &~{\rm TeV }, \\
0.46\pm 0.85 ~{\rm fb} 	& ~~~{\rm ATLAS} &{\rm at} ~\sqrt{s} = 8 &~{\rm TeV},  \\
5.6\pm2.4   ~{\rm fb} 	& ~~~{\rm CMS}    &{\rm at} ~\sqrt{s} = 13 &~{\rm TeV }, \\
6.2^{+2.4}_{-2.0} ~{\rm fb}  	& ~~~{\rm ATLAS} &{\rm at} ~\sqrt{s} = 13 &~{\rm TeV},  \\
\end{array}\right.
\end{eqnarray}
and the $\chi^2$ function we use is given by \cite{PDG}
\begin{eqnarray}
\chi^2 = \sum_{i=1}^4 \chi^2_i,
\end{eqnarray}
where
\begin{eqnarray}
\chi^2_i =\left\{\begin{array}{ll}
2[\mu_i^{exp} - \mu_i + \mu_i {\rm ln} \frac{\mu_i}{\mu_i^{exp}} ] 	& ~~~{\rm for~ the~ 13~TeV~ATLAS~ data}, \\
\frac{(\mu_i^{exp}-\mu_i)^2}{\sigma_{\mu_i^{exp}}^2}	& ~~~{\rm for~the~other~three~sets~of~data}, \\
\end{array}\right.
\end{eqnarray}
and $\mu_i$ is the theoretical prediction of the diphoton rate.

Beside the constraints from the different signals of $\phi$, the masses of the colored scalars are also constrained by the direct search for new particles
at the LHC. So far the most pertinent analysis for the $S_i$ pair production is to look for paired dijet resonances, which was performed by CMS
at the $8 \ {\rm TeV}$ LHC \cite{2-dijet-CMS}. This analysis concentrated on the process $ g g \to C C \to (jj) (jj)$ with $C$ denoting a color-octet vector
boson called coloron \cite{Coloron-model}, and it set the upper bounds on the cross section of the paired dijet events as a function of the coloron mass,
which were presented in Fig.7 of \cite{2-dijet-CMS}. In order to apply this analysis to our work, we first assume that the processes
$g g \to S_R S_R, S_I S_I \to 4 j$ have same cut efficiencies as those of the coloron pair production process in the analysis, then we calculate the $S_i$ production rates
at tree level to compare with the Fig.7 in \cite{2-dijet-CMS}.
We conclude that $m_S \gtrsim 450 {\rm GeV}$ can not be excluded.  We also note that the colored scalars may form bound states $O^0_+$, $O^0_R$ and $O^0_I$,
which were collectively called octetonia in \cite{octetonia}. The masses of these bound states are around $2 m_S$ and they can be produced directly
at the LHC to generate various signals such as $ g g \to O^0_i \to g g$, $W W^\ast$, $Z Z$, $\gamma \gamma$, $Z \gamma$  \cite{octetonia}.
In order to determine the mass bound imposed by the octetonia production, we calculate the rates of the signals at the $8\ {\rm TeV}$ LHC
by the formula presented in the appendix of \cite{octetonia}, and compare them with the corresponding LHC bounds introduced in this section.
We find that due to their relatively low
production rates, the octetonias as light as $750 {\rm GeV}$ are still experimentally allowed.
So in summary, it is fair to say that the colored scalars heavier than about $500 \ {\rm GeV}$ are still compatible with the LHC data.

\section{Numerical results and discussions} \label{result}

In order to answer whether the extended Manohar-Wise model can explain the diphoton data collected at the LHC
after considering the constraints described in Sec.\ref{const}, we fix $f=1 {\rm TeV}$ and $m_{S}= 600 {\rm GeV}$
($m_{S} = 1 {\rm TeV}$ as an alternative choice), and scan the following parameter space of the model
\begin{eqnarray}
 0 < \kappa \leq 10, \quad \quad |{\rm sin}\theta | \leq 0.05.
\end{eqnarray}
For each parameter point encountered in the scan, we first check whether it survives the constraints listed in Table I, then for the surviving
point we perform a fit to the diphoton data.

\begin{figure}[t]
  \includegraphics[width=14cm]{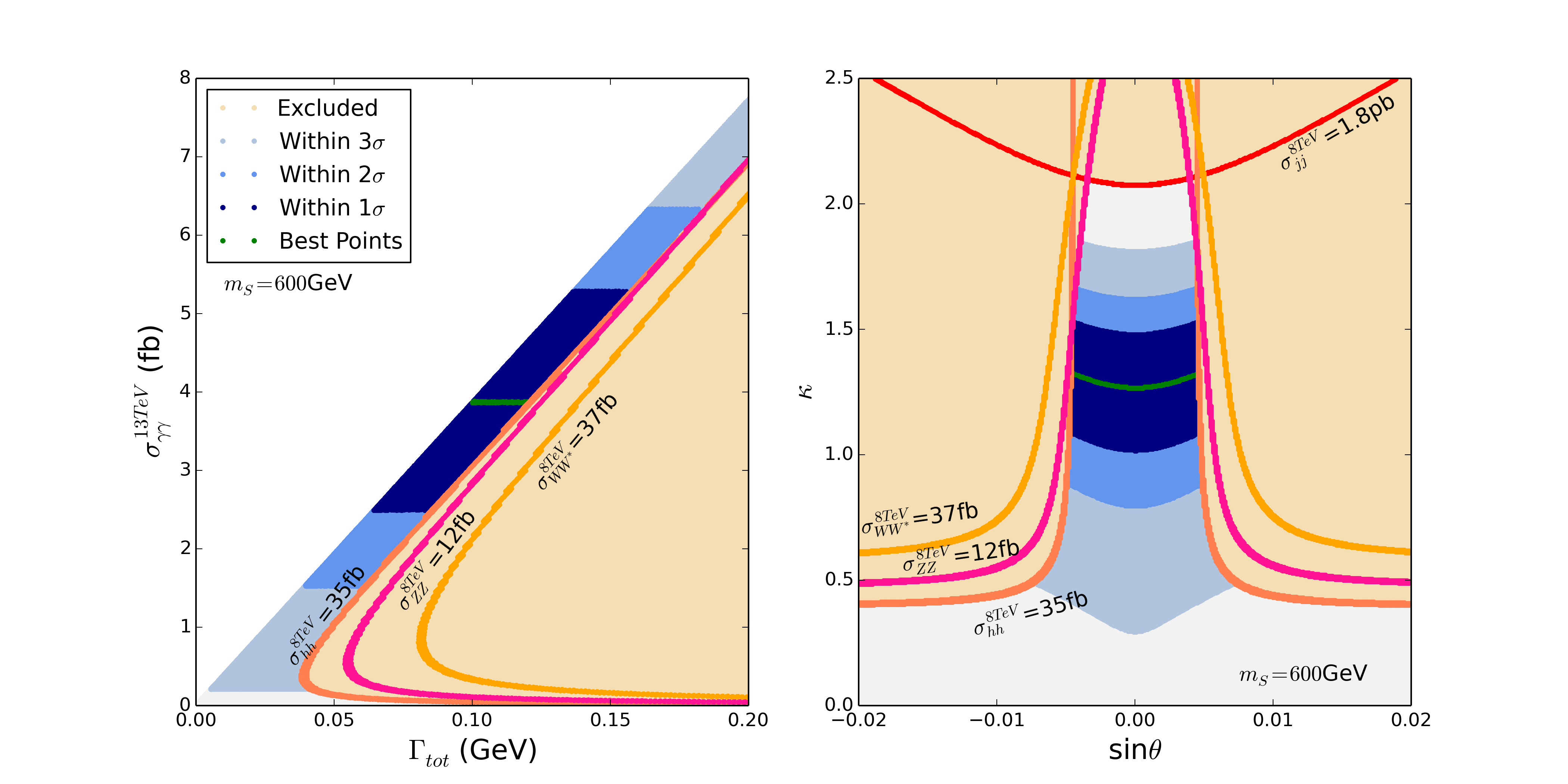}
  \includegraphics[width=14cm]{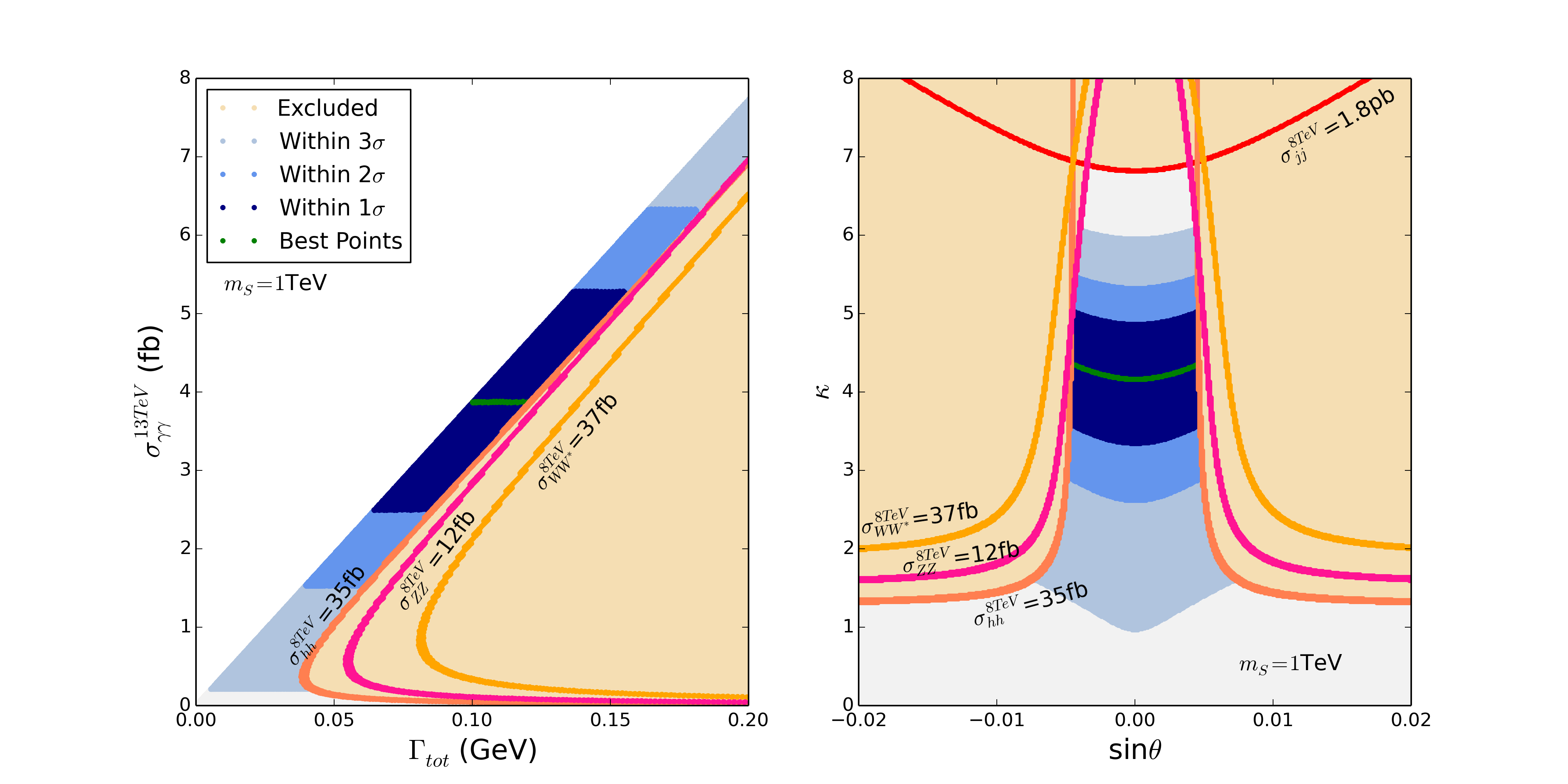}
  \caption{The fit results of the singlet extended Manohar-Wise model to the diphoton data together with the LHC Run I constraints listed in Table \ref{tab1}, which are projected on $\sigma_{\gamma \gamma}^{13 TeV}-\Gamma_{tot}$ and $\kappa-\sin \theta$ planes for $m_{S}=600 {\rm GeV}$ (upper panels) and  $m_{S}=1 {\rm TeV}$ (lower panels) respectively. The regions filled by the colors from gray to deep blue represent the parameter space that can fit the diphoton data within 3$\sigma$, 2$\sigma$ and 1$\sigma$ levels respectively, and by contrast the regions covered by straw color are excluded by the constraints.  The boundary lines from some signal channels at the LHC Run I are also plotted, and
  the green lines represent the parameter points which can predict the central value of the diphoton excess. }
  \label{fig2}
\end{figure}

Our results are showed in Fig.\ref{fig2} on the $\sigma_{\gamma \gamma}^{13 TeV}-\Gamma_{tot}$
planes (left panels) and $\kappa$-$\sin \theta$ planes (right panels) with the upper panels being for the $m_{S}=600 {\rm GeV}$ case and the lower panels for
the $m_{S}=1 \ {\rm TeV}$ case.  The regions filled by the colors from gray to deep blue represent the parameter space that survives the constraints and meanwhile is able to explain the diphoton data at 3$\sigma$, 2$\sigma$ and 1$\sigma$ levels respectively. In comparison, the regions covered by straw color are excluded by the constraints presented in Table \ref{tab1}.  The boundary lines
for some constraints in Table \ref{tab1} are also plotted with the right or upper side of the curves  being excluded for the left panels,
and the upper or outboard side of the curves being excluded for the right panels. The green lines represent the parameter points
which can predict the central value of the diphoton excess (about $3.9\  {\rm fb}$ in our fit). For these points, they predict
$\chi^2_{min} = 2.32$, which corresponds to a $p$-value of 0.68.

\begin{table}[t]
\small
\centering
\caption{The detailed information for two of the best points in our fits. }
\label{tab2}
\begin{tabular}{cccccccccccccc}
\hline
\hline
$\chi^2$   ~&~ $m_{S}$ ~&~  sin$\theta$ ~ &~ $\kappa$  ~&~  $\frac{\Gamma_{\phi\to gg}}{\Gamma_{H\to gg}^{SM}}$ ~&~ BR$_{\phi\to \gamma\gamma}$ ~&~  BR$_{\phi\to gg}$ ~&~ BR$_{\phi\to ZZ}$ ~&~ BR$_{\phi\to WW}$ & BR$_{\phi\to hh}$\\
\hline
2.32          ~&~ 600 GeV ~&~  -0.0023       ~&~ 1.30    ~&~ 4.46 ~&~ 0.12\% ~&~ 96.09\%    ~&~  0.62\%  ~&~ 1.57\%    ~&~1.35\% \\
2.32          ~&~ 1 TeV      ~&~  0.004       ~&~  4.31   ~&~  4.66 ~&~ 0.11\% ~&~  91.86\%    ~&~ 1.14\%  ~&~ 2.62\%    ~&~3.75\% \\
\hline
\end{tabular}
\end{table}

From Fig.\ref{fig2}, one can learn following facts
\begin{itemize}
  \item The singlet extension of the Manohar-Wise model can interpret the 750 GeV diphoton excess at $1\sigma$ level
    without conflicting with any constraints from the LHC Run I, and the corresponding parameter space is characterized by
    $|\sin \theta| < 0.01$ and $\kappa f > 1 {\rm TeV}$. In Table \ref{tab2}, we list the detailed information for two of
    the best points in our fits with one corresponding to the $m_S=600$ GeV case
     and the other corresponding to the $m_S=1 \ {\rm TeV}$ case.
  \item For the degenerated colored scalars, the choice of their masses does not influence the fit
       quality. This is obvious by comparing the left upper panel with the left lower panel.
       In fact, if a moderately higher value of $m_{S}$ is adopted, one can always choose a larger $\kappa f$
       to keep $\Gamma_{\phi \to g g}$ and  $\Gamma_{\phi \to \gamma \gamma}$ roughly unchanged (see the formulae in Section III).
       This feature has been reflected in Table \ref{tab2} for the two best points.
  \item Since $\sin \theta$ is small in order to satisfy the constraints in Table \ref{tab1}, $\phi \to g g$ is always the dominant decay mode, and
        consequently the total width of $\phi$ is less than about $0.2 \ {\rm GeV}$. We point out that, if the theory is embedded in a more complex framework,
        in principle $\phi$ may decay dominantly into other particles such as the dark matter candidate and/or multi-jets so that its width can be
        enhanced greatly. Considering $\sigma_{\gamma \gamma}^{13 TeV} \propto (\kappa f)^4/\Gamma_{tot}$, one can infer that if $\Gamma_{tot}$ is enlarged
        by a factor of 100, $\kappa f$ has to be enhanced by roughly $3.2$ times to keep $\sigma_{\gamma \gamma}^{13 TeV}$ unchanged.
        This situation is acceptable if the colored scalars are moderately light (so that $\kappa f$ can be relatively small). So our explanation for the excess
        is still viable even for a significantly wider width of $\phi$.
  \item The signal channels listed in Table \ref{tab1}, especially the $hh$ and $ZZ$ channels, have tightly limited the parameter space of
        the model to explain the diphoton excess. For example, the resonant $hh$ signal requires $\sin \theta \lesssim 0.01$, and the dijet signal
        imposes an upper bound on $\kappa f$.

\end{itemize}

\section{Other aspects of our explanation }\label{vacuum-stability}

\begin{figure}[t]
  \includegraphics[width=15.5cm]{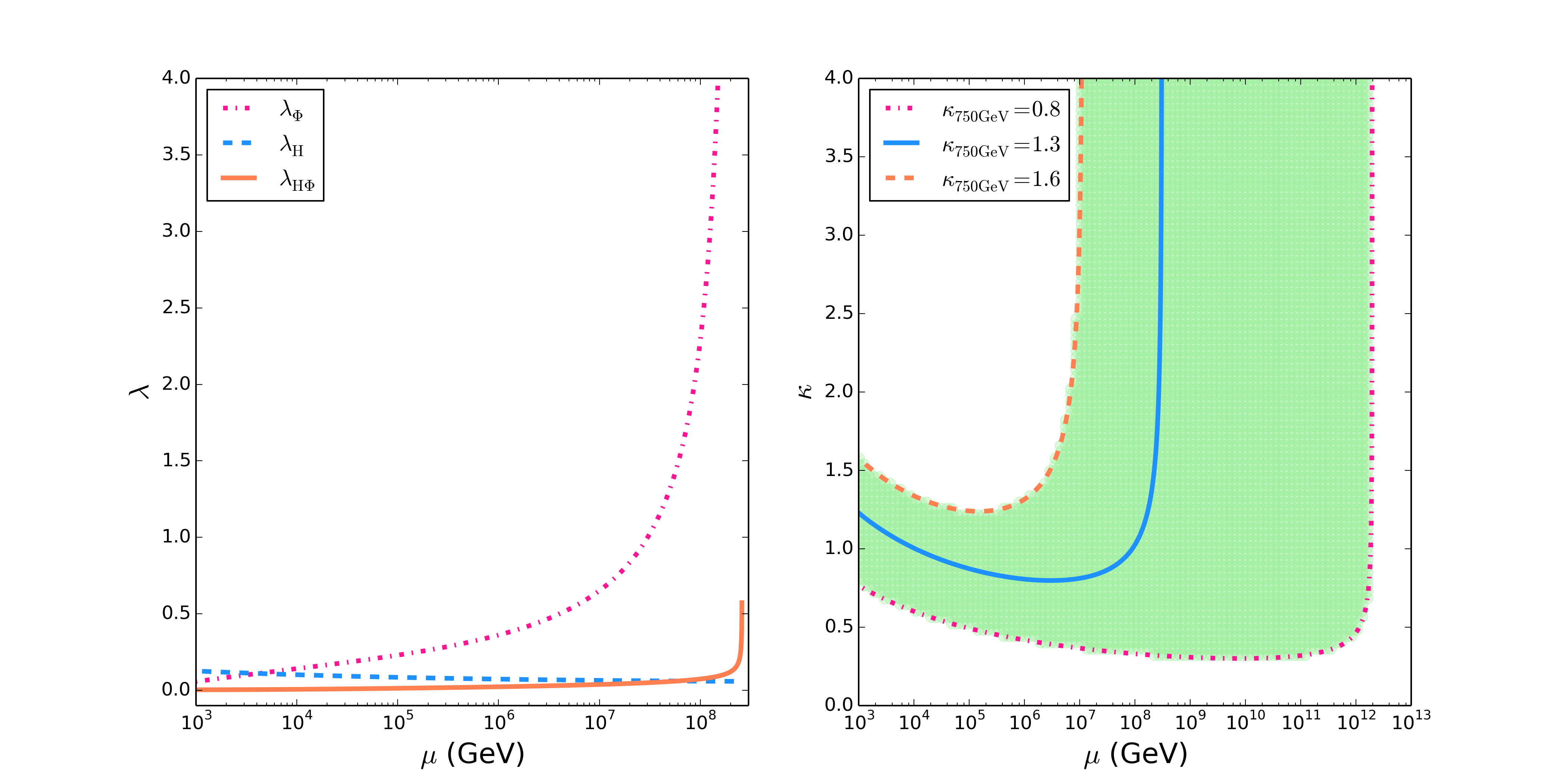}
  \caption{Left panel: the dependence of the coefficients $\lambda_\Phi$, $\lambda_H$ and $\lambda_{H\Phi}$ on the renormalization group equation
  scale $\mu$ for the first benchmark
  point in Table \ref{tab2}. Right panel: the evolutions of the coefficient $\kappa$ with the energy scale. In getting these panels,
  we have set $\sin \theta = -0.0023$, $m_S = 600 {\rm GeV}$ and $f= 1 {\rm TeV}$, then $\kappa_{750 {\rm GeV}} =1.3$ corresponds to the benchmark point,
  and $\kappa_{750 {\rm GeV}} = 0.8, 1.6$ are the boundary values of $\kappa$ to explain the diphoton excess at $2 \sigma$ level. }
  \label{fig3}
\end{figure}

From the discussions in above section, one can learn that our explanation of the diphoton excess is featured by a moderately large $\kappa$ for a fixed $f$.
In this case, the evolutions of $\kappa$, $\lambda_{\Phi}$ and $\lambda_{H\Phi}$ with the energy scale by the renormalization group
equation (RGE) can change greatly their values at the scale $m_\phi = 750 {\rm GeV}$, which is obvious from the $\beta$ functions
of the coupling coefficients listed in the appendix.  As a result,
the vacuum may become unstable, and/or the theory may become non-perturbative when the energy scale surpasses a critical value.
In the following, we address these issues.

For large field values $H^0, \Phi \gg v, f$, the potential of the color-singlet scalars is very well approximated by its $RG$-improved tree-level expression \cite{vacuum-SM},
\begin{eqnarray}
V_{eff}^{tree} \simeq \lambda_{\Phi} (\mu) \Phi^4 + \frac{\lambda_H (\mu)}{4} (H^0)^4 + \frac{\lambda_{H \Phi}}{2} \Phi^2 (H^0)^2,
\end{eqnarray}
where $\mu$ is the scale of the RGE, and it satisfies $\mu \gg m_\phi, f$ in our interested case. Then the condition of the absolute
stability of the potential is given by
\begin{eqnarray}
4 \lambda_H (\mu) \lambda_\Phi (\mu) -  \lambda_{H \Phi}^2 (\mu) > 0.
\end{eqnarray}
In order to check wether this condition is satisfied,  we show in the left panel of  Fig.\ref{fig3} the dependence of $\lambda_\Phi$,
$\lambda_H$ and $\lambda_{H\Phi}$ on the energy scale for the first benchmark point in Table \ref{tab2}. This panel indicates that, due to the push up of the
$\kappa^2$ term in the $\beta_{\lambda_\Phi}$ function, $\lambda_\Phi$ increases rapidly with $\mu$, and consequently the
stability condition is always satisfied. But on the other hand, we note from the panel that $\lambda_\Phi$ will reach its Landau pole at $\mu \simeq 2 \times 10^8 {\rm GeV}$.
Considering that the $\beta_\kappa$ function also contains the $\kappa^2$ term, we imagine that $\kappa$ should also have such a behavior.
The evolution of $\kappa$ with the RGE scale is presented in the right panel of Fig.\ref{fig3} for different choices of $\kappa$.
Here $\kappa_{750 {\rm GeV}} = 1.3$ corresponds to the first benchmark point in Table
\ref{tab2}, and $\kappa_{750 {\rm GeV}} = 0.8, 1.6$ are the boundary values of $\kappa$ to explain the diphoton excess at $2 \sigma$ level
if $\sin \theta$, $m_S$ and $f$ take the same values as the benchmark point.  This panel indicates that, for the parameter region considered to explain the excess,
our theory keeps perturbative until the scale at about $10^7 {\rm GeV}$. Given that this scale is much higher
than the electroweak scale, we think that our explanation of the excess is acceptable.

Before we end the discussion, we emphasize that the diphoton excess actually imposes constraints on the product $\kappa f$.  So if one chooses a larger value of $f$,
a lower value of $\kappa$ is still capable in explaining the excess. This will postpone the appearance of the Landau poles. For example, if
$f= 1 {\rm TeV}$ for the first point in Table \ref{tab2} is switched to $f= 2 {\rm TeV}$  so that the $\kappa$ is decreased by one half,
we checked that the Landau pole will appear at about $2 \times 10^{9} {\rm GeV}$, which is about one order higher than the $f = 1 {\rm TeV}$ case.

\section{Conclusion}\label{sum}

The evidence of a new scalar particle $X$ from the 750 GeV diphoton excess, and the absence of any other signal of new physics at the LHC so far
suggest the existence of new colored scalars, which may be moderately light and thus can induce sizable $X g g$ and $X \gamma \gamma$ couplings without
resorting to very strong interactions. Motivated by this speculation, we extended the Manohar-Wise model by adding one gauge singlet scalar field.
The resulting theory then predicts one singlet dominated scalar $\phi$
as well as three kinds of colored scalars, which can mediate through loops the $\phi gg$ and $\phi \gamma \gamma$ interactions. Within this theoretical framework,
we investigated whether the diphoton excess can be explained by the process $ g g \to \phi \to \gamma \gamma$.
For this purpose, we scanned the parameter space of the model by considering the constraints from various LHC Run I signals, and for each surviving sample,
we performed a fit to the $750$ GeV diphoton data collected at both the 8 TeV and the 14 TeV LHC.  It turns out that in reasonable parameter regions
of the model, the diphoton excess can be explained at $1 \sigma$ level without conflicting with any experimental
constraints. The best points in the fit can predict the central value of the excess with $\chi_{min}^2=2.32$, which corresponds
to a $p$-value of $0.68$.

\appendix

\section{$\beta$ functions for various coupling coefficients}

\vspace{-0.4cm}

In this appendix, we investigate the $\beta$ functions of different coupling coefficients in the singlet extended Manohar-Wise model.
Compared with the original version of the model, the extended theory introduces new operators such as $\Phi^4$, $\Phi^2 |H|^2$ and
$\Phi^2 Tr(S^{\dag j} S_j)$. As a result, some $\beta$ functions for $\lambda_i$ get additional contributions, and meanwhile
new $\beta$ functions corresponding to the operators appear.  In this work, we get the expressions of these functions by
the technique in \cite{RGE}, and verify them by comparing parts of our results with their corresponding ones in \cite{Beta-function}
where the $\beta$ functions for $\lambda_i$ were obtained in the Manohar-Wise model and also with
those in \cite{octet-vacuum} where all $\beta$ functions in the singlet extension of the SM were given. In the following,
we only list the $\beta$ functions that are affected by $\kappa$, $\lambda_{H \Phi}$ and $\lambda_\Phi$, which are given by
\begin{eqnarray}
(16 \pi^2) \beta_\kappa  &=& 4 \kappa^2 + 24 \kappa \lambda_{\Phi} + 17 \kappa \lambda_8 + 4 \lambda_1 \lambda_{H \Phi} + 2 \lambda_2 \lambda_{H \Phi}
- ( 18 g_3^2 + \frac{9}{2} g_2^2 + \frac{9}{10} g_1^2) \kappa, \\
(16 \pi^2) \beta_{\lambda_H} &=& 24 \lambda_H^2 + 2 \lambda_{H\Phi}^2 + 4 \lambda_1^2 + 2 \lambda_2^2 + 4 \lambda_1 \lambda_2 + 4 |\lambda_3|^2 \nonumber \\
& & - (9g_2^2 + \frac{9}{5} g_1^2)\lambda_H + 12 y_t^2 \lambda_H - 6y_t^4 +  \frac{1}{2}\ (\frac{27}{100} g_1^4 + \frac{9}{10} g_1^2 g_2^2 + \frac{9}{4}g_2^4),  \\
(16 \pi^2) \beta_{\lambda_{\Phi}} &=& 72 \lambda_{\Phi}^2 + 2 \lambda_{H\Phi}^2 + 4 \kappa^2, \\
(16 \pi^2) \beta_{\lambda_{H\Phi}} &=& 8 \lambda_{H \Phi}^2 + 12 \lambda_{H\Phi} \lambda_{H} + 24 \lambda_{H\Phi}\lambda_{\Phi} + 8 \kappa \lambda_1 + 4 \kappa \lambda_2 \nonumber \\
 & & -(\frac{9}{2} g_2^2 + \frac{9}{10} g_1^2)\lambda_{H\Phi} +6 y_t^2 \lambda_{H\Phi}, \\
(16 \pi^2) \beta_{\lambda_1} &=& 2 \lambda_1^2 + 17 \lambda_1 \lambda_8  + 12 \lambda_1 \lambda_H  + 4 \lambda_2 \lambda_H + 4 \kappa \lambda_{H \Phi} \nonumber \\
 && - (18g_3^2 + 9 g_2^2 + \frac{9}{5} g_1^2 )\lambda_1 + 6 y_t^2 \lambda_1 + 6 (\frac{27}{100} g_1^4 + \frac{9}{10}g_1^2 g_2^2 +\frac{9}{4} g_2^4) + \beta_{\lambda_1}^\prime, \\
(16 \pi^2) \beta_{\lambda_8} &= & 20 \lambda_8^2 + 2 \lambda_{1}^2 + 2 \kappa^2 - (36 g_3^2 + 9 g_2^2 + \frac{9}{5}g_1^2) \lambda_8 \nonumber \\
 & & + 2 (\frac{27}{100} g_1^4 + \frac{9}{10} g_1^2 g_2^2 + \frac{9}{4} g_2^4 ) + \beta^\prime_{\lambda_8},
\end{eqnarray}
where
\begin{eqnarray}
(16 \pi^2) \beta_{g_i} &= & b_i g_i^3,\quad \quad i=1,2,3,  \\
(16 \pi^2) \beta_{y_t} &= & y_t (\frac{9}{2} y_t^2 - \frac{17}{20} g_1^2 - \frac{9}{4} g_2^2 - 8 g_3^2 ),
\end{eqnarray}
with $(b_1,b_2,b_3)=(41/10,-19/6,-7)$ in the SM, and $(49/10,-11/6,-5)$ after considering the effect of the colored scalars, and $\beta^\prime_{\lambda_1}$ and $\beta^\prime_{\lambda_8}$ denote
the effects from the couplings $\lambda_{2-7}$ and $\lambda_{9-11}$ with their lengthy expressions presented
in \cite{Beta-function}. Note that in getting these expressions, we have used the standard normalization $g_1^2=\frac{5}{3} g_Y^2$. Also note that these
$\beta$ functions are valid only at the energy scale where the colored scalars are active.

Finally, we reminder that the expressions of the $\beta$ functions not listed here, which are for $\lambda_{2-7}$ and $\lambda_{9-11}$, were given in the
appendix of \cite{Beta-function}. These functions are not important for our calculation since they are unaffected directly
by the large coefficient $\kappa$, and meanwhile the initial values of $\lambda_{1-11}$ at the scale $m_\phi = 750 {\rm GeV}$
are set to be very small. In fact, we checked that the effects of $\lambda_{2-7}$ and $\lambda_{9-11}$ on the running of $\kappa$, $\lambda_\Phi$,
$\lambda_{H\Phi}$, $\lambda_{H}$, $\lambda_1$ and $\lambda_8$ are negligibly small.

\vspace{1cm}

\noindent {\bf{Acknowledgements}}
We would like to thank Fei Wang, Peihua Wan and Guanghua Duan for helpful discussions and comments. This work was supported by the National
Natural Science Foundation of China (NNSFC) under grant No. 10821504, 11222548, 11121064, 11135003, 90103013 and 11275245,
the CAS Center for Excellence in Particle Physics (CCEPP) and World Premier International Research Center Initiative (WPI Initiative), MEXT, Japan.

\end{document}